\documentclass[prl,aps,twocolumn]{revtex4}

\usepackage{latexsym}
\usepackage{amssymb}
\usepackage{amsmath}
\usepackage[varg]{txfonts}
\usepackage{mathrsfs}
\usepackage{upgreek}
\usepackage[small]{caption}
\usepackage[]{subfig}
\usepackage{verbatim}
\usepackage{array}
\usepackage{color}
\usepackage{graphicx}
\usepackage{slashbox}
\DeclareMathAlphabet{\mathpzc}{OT1}{pzc}{m}{it}
\usepackage{hyperref}

\begin{document}
\title{Direct observation of the glue pairing the halo of the nucleus $^{11}$Li}
\author{G. Potel}
\affiliation{Dipartimento di Fisica, Universit\`{a} di Milano,
Via Celoria 16, 20133 Milano, Italy.}
\affiliation{INFN, Sezione di Milano Via Celoria 16, 20133 Milano, Italy.}
\author{F. Barranco}
\affiliation{Departamento de Fisica Aplicada III, Universidad de Sevilla, Escuela Superior de Ingenieros,
Sevilla, 41092 Camino de los Descubrimientos s/n,
Spain.}
\author{E. Vigezzi}
\affiliation{INFN, Sezione di Milano Via Celoria 16, 20133 Milano, Italy.}
\author{R. A. Broglia}
\affiliation{Dipartimento di Fisica, Universit\`{a} di Milano,
Via Celoria 16, 20133 Milano, Italy.}
\affiliation{INFN, Sezione di Milano Via Celoria 16, 20133 Milano, Italy.}
\affiliation{The Niels Bohr Institute, University of Copenhagen, Blegdamsvej 17,
2100 Copenhagen {\O}, Denmark.}
\begin{abstract}
With the help of a unified nuclear structure--direct reaction theory we analyze the reaction $^1$H($^{11}$Li,$^{9}$Li)$^3$H. The two halo neutrons are correlated through the bare and the induced (medium polarization) pairing interaction. By considering all dominant reaction processes leading to the population of the $1/2^-$ (2.69 MeV) first excited state of $^9$Li, namely multistep transfer (successive, simultaneous and non--orthogonality), breakup and inelastic channels, it is possible to show that the experiment provides, for the first time in nuclear physics, direct evidence of phonon mediated pairing.

\end{abstract}
\
\maketitle
There exists conspicuous circumstantial evidence which testifies to the important role medium polarization effects play in the phenomenon of nuclear superfluidity (see e.g. \cite{Brink:05} and refs. therein). In spite of this, a quantitative assessment of it is still lacking.
 Specially promising in this quest are highly polarizable exotic nuclei, in particular, the light halo nucleus $^{11}$Li, for which, the balance between bare and induced pairing interactions is strongly shifted in favour of the induced interaction
  (\cite{Barranco:01}, see also \cite{Hagino:07}, \cite{Masui:05}, \cite{Myo:08}, \cite{Ayoama:02}, \cite{VinhMau:96}).

 In this nucleus, the last two neutrons are very weakly bound ($S_{2n}\approx 380$keV \cite{Bachelet:08}, \cite{Smith:08}, \cite{Roger:09}). If one neutron is taken away from $^{11}$Li, a second neutron will come out immediately leaving behind the core of the system, the ordinary nucleus $^{9}$Li. This result testifies to the fact that pairing is central in the stability of $^{11}$Li (see e.g. \cite{Nunes:05}, \cite{Hagino:05}).

In ref. \cite{Barranco:01} it has been shown
 that the two outer (halo) neutrons of $^{11}$Li in its ground state attract each other, not only due to the strong nuclear force acting among them, but also and primarily due to the virtual processes associated with the exchange of collective vibrations. In particular, the quadrupole vibration of the  $^{9}$Li core, and the dipole vibration associated with the neutron halo field (pigmy resonance of  $^{11}$Li \cite{Nakamura:06}). Such a pairing mechanism is clearly reflected in the calculated ground state wavefunction of  $^{11}$Li \cite{Barranco:01},
\begin{equation}\label{eq1}
    |^{11}\text{Li}(gs);3/2^-\rangle = |\tilde 0\rangle_\nu \otimes |1 p_{3/2}(\pi)\rangle,
\end{equation}
where $\pi$ and $\nu$ indicate proton and neutron degrees of freedom respectively, while $|\tilde 0\rangle_\nu$ indicates the halo neutron Cooper pair wavefunction, that is,
\begin{equation}\label{eq2}
    |\tilde 0\rangle_\nu = | 0\rangle +  \alpha |(p_{1/2},s_{1/2})_{1^-}\otimes 1^-;0\rangle+\beta|(s_{1/2},d_{5/2})_{2^+}\otimes 2^+;0\rangle,
\end{equation}
with
\begin{equation}\label{eq5}
 \alpha\approx 0.7, \quad \text{and} \quad \beta\approx 0.1,
\end{equation}
and
\begin{equation}\label{eq4}
    | 0\rangle=0.45 |s_{1/2}^2(0)\rangle + 0.55 |p_{1/2}^2(0)\rangle+0.04|d_{5/2}^2(0)\rangle,
\end{equation}
the states $| 1^-\rangle$ and $| 2^+\rangle$ being the (RPA) states describing the dipole pigmy resonance of $^{11}$Li and  the quadrupole vibration of the core $^9$Li (see \cite{Barranco:01}, see also Tables 11.3 and 11.5 of ref \cite{Brink:05}).
 The intrinsic non--observability of virtual processes (like the exchange of collective vibrations between Cooper pair partners leading to the second and third components of the state $|\tilde 0\rangle_\nu$) is a fact. However, in those cases in which the experimental tool exists which specifically probes the phenomenon under study, one can force the virtual processes of interest to become real. In this way one could, for example, hope to observe the collective vibrations of $^{11}$Li and of  $^{9}$Li correlating the two--halo neutrons, with the help of a two--particle transfer process, specific probe of pairing in nuclei (\cite{Bohr:76}, \cite{Broglia:73}). In what follows we shall show, with the help of a quantitative reaction plus nuclear structure analysis, that the experiment $^1$H($^{11}$Li,$^{9}$Li)$^3$H recently carried out at TRIUMF (\cite{Tanihata:08}), provides direct evidence of phonon exchange between nuclear Cooper pair partners.

\begin{figure*}
\centerline{\includegraphics*[width=.8\textwidth,angle=0]{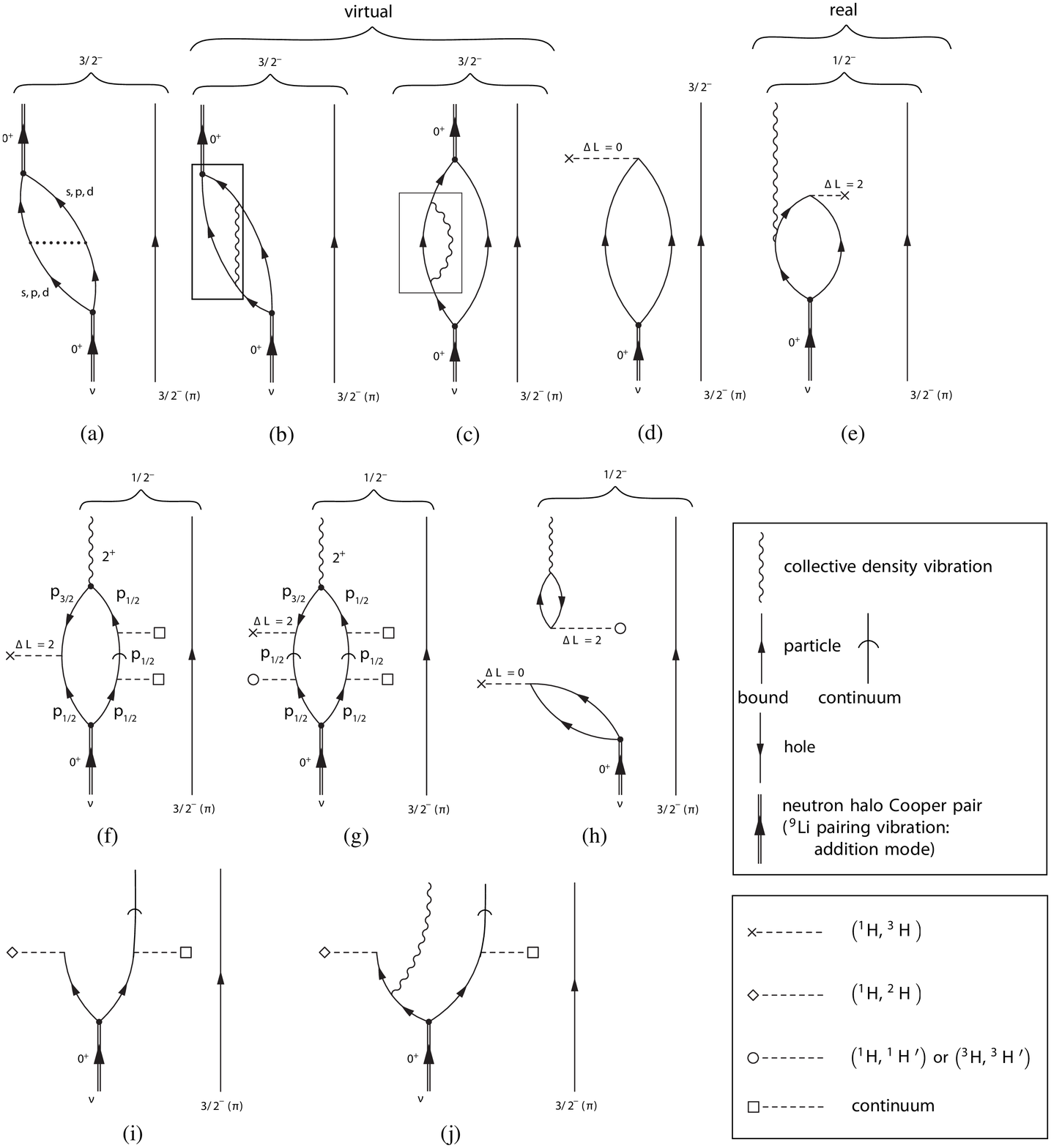}}
\caption{\protect\footnotesize Representative Nuclear Field Theory--Feynman diagrams associated with correlation process ((a),(b),(c)) and with one-- and two--particle pick--up reactions ((i),(j) and (d),(e) respectively) of the halo neutrons of $^{11}$Li (Cooper pair, indicated in terms of a double arrowed line). Also shown are the possible diagrams associated with other channels (breakup and inelastic) populating the $1/2^-$ (2.69 MeV) state: f) one of the neutrons is picked up (the other one going into the continuum, i.e. breaking up from the $^9$Li core) together with a neutron from the $p_{3/2}$ orbital of the $^9$Li core  leading, for those events in which the neutron moving in the continuum falls into the $p_{3/2}$ hole state, leading to the excitation of the $1/2^-$ final state ($2^+$ density mode (wavy line)) coupled to the $p_{3/2}(\pi)$. g) the proton field acting once breaks the Cooper pair forcing one of the halo neutrons to populate a $p_{1/2}$ continuum state (the other one follows suit), while acting for the second time picks up one of the neutrons moving in the continuum and another one from those moving in the $p_{3/2}$ orbital of $^9$Li eventually leaving the core in the quadrupole mode of excitation. In (h) the  two--step transfer plus inelastic final channel process exciting the $(2^+\otimes p_{3/2}(\pi))_{1/2^-}$ state is shown.}\label{fig1}
\end{figure*}

To convey the details of such an analysis, which is based on the nuclear structure description of $^{11}$Li reported in \cite{Barranco:01}, Nuclear Field Theory (NFT)--Feynman diagrams (see e.g. refs. \cite{Bes:76c}, \cite{Mottelson:76}, \citep{Bortignon:77} and their generalization to deal with reaction processes \cite{Broglia:05c}) are used (Fig.\ref{fig1}).
In reference \cite{Barranco:01}, the two halo neutrons correlate through the bare interaction (Fig. \ref{fig1}(a)) and through the exchange of collective vibrations, leading to self--energy (see \cite{Mahaux:85} and refs. therein) and vertex corrections (boxed processes in Figs. \ref{fig1}(c) and \ref{fig1}(b) respectively; see also Eqs. (\ref{eq1})--(\ref{eq4})). Solving the associated eigenvalue problem a bound Cooper pair is obtained ($S_{2n}$=330 keV).

 From the diagrams displayed in Figs. \ref{fig1} (b) and \ref{fig1} (c), it is easy to understand how the virtual propagation of collective vibrations (in the present case $1^-$ and $2^+$ vibrations) can be forced to become a real process: by transferring one or two units of angular momentum in a two--neutron pick up process. In particular, the correlation mechanism displayed in Figs. \ref{fig1} (b) and \ref{fig1} (c) predict a direct excitation of the quadrupole multiplet of $^9$Li (see Fig. \ref{fig1}(e), see also \cite{Brink:05} Fig. 11.6). On the other hand, if the two--neutron pick--up process takes place before the virtual excitation of the vibrational mode, the ground state of $^9$Li is populated (Fig. \ref{fig1} (d)).

Of course the $1/2^-$ (2.69 MeV) first excited state of $^9$Li can also be excited through a break up process in which one (see Fig. \ref{fig1}(f)), or both neutrons (see Fig. \ref{fig1}(g)) are forced into the continuum for then eventually one of them to fall into the $1p_{3/2}$ orbital of $^8$Li and excite the quadrupole vibration of the core, in keeping with the fact that the main RPA amplitude of this state is precisely $X(1p^{-1}_{3/2},1p_{1/2})(\approx 1)$ (cf. ref \cite{Barranco:01}). The remaining channel populating the first excited state of $^9$Li is associated with an inelastic process (see Fig. \ref{fig1}(h)): two--particle transfer to the ground state of $^{9}$Li and Final State (inelastic scattering) Interaction (FSI) between the outgoing triton and $^{9}$Li in its ground state, resulting in the inelastic excitation of  the $1/2^-$ state.

Making use of the wavefunctions of reference \cite{Barranco:01} and of software developed on purpose to take into account microscopically all the different processes mentioned above, that is 9 different reaction channels and continuum states up to 50 MeV of excitation energy, we have calculated the corresponding transfer amplitude and associated probabilities $p_l$, a technically demanding calculation which can be considered a proper \textit{tour de force}.

 In Table \ref{tab1} we display the probabilities $p_l=|S_l^{(c)}|^2$ associated with each of the processes discussed above, where the amplitude $S_l^{(c)}$ is related to the total cross section associated with each of the channels $c$  by the expression \cite{Satchler:80}, \cite{Landau:81}
\begin{equation}\label{eq6}
    \sigma_c=\frac{\pi}{k^2}\sum_l(2l+1)|S_l^{(c)}|^2,
\end{equation}
$k$ being the wave number of the relative motion between the reacting nuclei.

 In keeping with the small values of $p_l$, in what follows we take into account the interference between the contributions associated with the different reaction paths making use of second order perturbation theory, instead of a coupled channel treatment \cite{Ascuitto:69} \cite{Tamura:70} \cite{Khoa:04} \cite{Keeley:07b}. In particular in the case of the $1/2^-$ (2.69 MeV) first excited state of $^9$Li,
\begin{equation}
    \frac{d\sigma}{d\Omega}(\theta)=\frac{\mu^2}{16\pi^3\hbar^4}\left|\sum_l(2l+1)P_l(\theta)\sum_{c=2}^5 T^{(c)}_l\right|^2,
\end{equation}
where $\mu$ is the reduced mass and $T^{(c)}_l$ are the transition matrix elements (in the DWBA \cite{Satchler:80}) associated with the different channels and for each partial wave.

\begin{table}
\begin{center}
\begin{tabular}{|c|c|c|c|c|c|}
\hline
\backslashbox {$l$}{$c$} & \textbf{1} & \textbf{2} & \textbf{3} & \textbf{4}& \textbf{5} \\
\hline
 0& $4.35\times 10^{-3}$ &$1.79\times 10^{-4}$ & $4.81\times 10^{-6}$& $2.90\times 10^{-11}$& $3.79\times 10^{-8}$\\
\hline
 1& $3.50\times 10^{-3}$& $9.31\times 10^{-4}$& $1.47\times 10^{-5}$&$1.87\times 10^{-9}$& $1.09\times 10^{-6}$\\
\hline
 2& $7.50 \times 10^{-4}$& $8.00\times 10^{-5}$& $2.45\times 10^{-5}$&$1.25\times 10^{-8}$&$1.21\times 10^{-6}$\\
\hline
 3& $6.12\times 10^{-4}$&$9.81\times 10^{-5}$ & $1.51\times 10^{-6}$&$6.50\times 10^{-10}$&$2.20\times 10^{-7}$\\
\hline
 4&$1.10\times 10^{-4}$ &$ 1.18\times 10^{-5}$ & $2.21\times 10^{-7}$&$4.80\times 10^{-11}$&$1.46\times 10^{-8}$ \\
\hline
 5& $3.65\times 10^{-5}$& $2.16\times 10^{-7}$& $7.42\times 10^{-9}$&$6.69\times 10^{-13}$&$9.63\times 10^{-10}$\\
\hline
 6& $1.35\times 10^{-5}$& $6.05\times 10^{-8}$&$2.88\times 10^{-10}$ &$8.04\times 10^{-15}$&$1.08\times 10^{-11}$\\
\hline
 7& $4.93\times 10^{-6}$& $7.78\times 10^{-8}$& $6.01\times 10^{-11}$&$4.05\times 10^{-16}$&$5.26\times 10^{-13}$\\
\hline
 8& $2.43\times 10^{-6}$& $2.62\times 10^{-8}$& $7.4\times 10^{-12}$&$1.26\times 10^{-17}$&$9.70\times 10^{-11}$\\
\hline
\end{tabular}
\caption{Probabilities $p_l$  associated with the processes described in the text for each partial wave $l$. The different channels are labeled by a channel number $c$ equal to: \textbf{1}, multistep transfer to the $^9$Li ground state (Fig. \ref{fig1}(d)); \textbf{2}, multistep transfer (Fig. \ref{fig1}(e)), \textbf{3}, breakup (Fig. \ref{fig1}(f)), \textbf{4}, breakup  (Fig. \ref{fig1}(g)), and \textbf{5} inelastic processes (Fig. \ref{fig1}(h)) involved in the population of the $1/2^-$ (2.69 MeV) first excited state of $^9$Li. Of notice that the probabilities displayed in columns \textbf{1} and \textbf{2} result from the (coherent) sum of three amplitudes namely those associated with successive, simultaneous and non--orthogonality transfer channels (see also Fig. \ref{fig3}).}\label{tab1}
\end{center}
\end{table}

\begin{figure}
\centerline{\includegraphics*[width=.4\textwidth,angle=0]{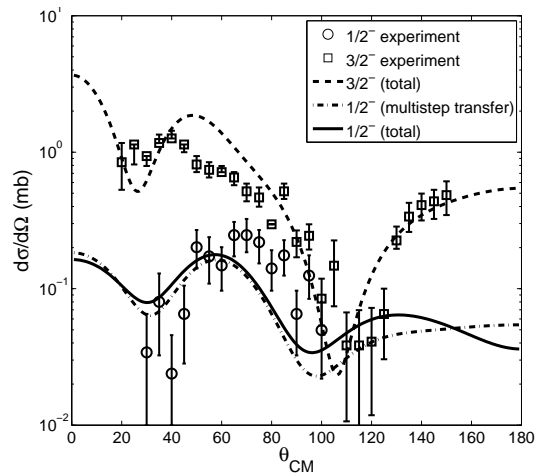}}
\caption{\protect\footnotesize Experimental (\cite{Tanihata:08}) and theoretical differential cross sections (including multistep transfer as well as breakup and inelastic channels).  of the
$^1$H($^{11}$Li,$^9$Li)$^3$H  reaction populating the ground state ($3/2^-$) and the first excited state ($1/2^-$; 2.69 MeV) of $^{9}$Li. Also shown (dash--dotted curve) is the differential cross section associated with this state but taking into account only multistep transfer. The optical potentials used are from \cite{Tanihata:08} and \cite{An:06}.}\label{fig2}
\end{figure}

\begin{figure}
\centerline{\includegraphics*[width=.7\textwidth,angle=0]{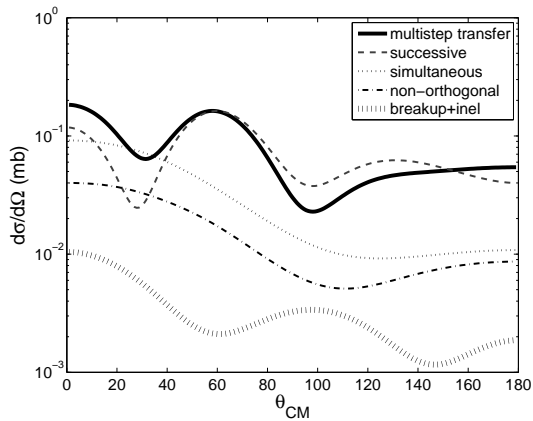}}
\caption{Successive, simultaneous and non-orthogonality contributions
to the  $^1$H($^{11}$Li,$^9$Li)$^3$H differential cross section
associated with the $1/2^-$ state
of $^9$Li, displayed in Fig. \ref{fig2}. Also shown is the (coherent) sum of the breakup ($c=3$ and 4) and inelastic ($c=5$) channel contributions.}\label{fig3}
\end{figure}

 Making use of all the elements discussed above, multistep transfer (see e.g.\cite{Bayman:82},\cite{Igarashi:91}, \cite{Bayman:73} as well as \cite{Broglia:05c}), breakup and inelastic channels were calculated, and the results displayed in Figs. \ref{fig2} and \ref{fig3} and in Table \ref{tab2}. Theory provides an overall account of the experimental findings. In particular, in connection with the $1/2^-$ state, this result essentially emerges from cancellations and coherence effects taking place between the three terms contributing to the multistep two--particle transfer cross section (see Fig. \ref{fig3}), tuned by the nuclear structure amplitudes associated with the process shown in Fig. \ref{fig1} (e) as well as Eqs. (\ref{eq1})--(\ref{eq4}). In fact, and
as shown in Figs. \ref{fig2} and \ref{fig3}, the contribution of inelastic and break up processes (Figs. \ref{fig1}(f),(g) and (h) respectively) to the population of the $1/2^-$ (2.69 MeV) first excited state of $^9$Li are negligible as compared with the process depicted in Fig. \ref{fig1}(e). In the case of the breakup channel this is a consequence of the low bombarding energy of the $^{11}$Li beam (inverse kinematics), combined with the small overlap between continuum (resonant) neutron $p_{1/2}$ wavefunctions and  bound state wavefunctions. In the case of the inelastic process (Fig. \ref{fig1}(h)), it is again a consequence of the relative low bombarding energy. In fact, the adiabaticity parameters $\xi_C,\xi_N$ (see eqs. (IV.12) and (IV.14) of ref. \cite{Broglia:05c}) associated with Coulomb excitation and inelastic excitation in the t+$^9$Li channel are larger than 1, implying an adiabatic cutoff. In other words, the quadrupole mode is essentially only polarized during the reaction but not excited. The situation is quite different in the case of the virtual process displayed in Fig. \ref{fig1} (e). Being this an off--the--energy shell process, energy is not conserved, and adiabaticity plays no role.

Of notice that the final states observed in the two neutron pick up process can, in principle, also be populated in a one--particle pick up process (see Figs. \ref{fig1}(i) and \ref{fig1}(j)). This prediction could likely be checked with the same experimental setup used in \cite{Tanihata:08}.
\begin{table}
\begin{center}
\begin{footnotesize}
\begin{tabular}{|c|c|c|c|}
\hline
& $\sigma$($^{11}$Li(gs)  $\to$ $^9$Li (i)) (mb) & & \\
\hline
i & $\Delta L$ & Theory & Experiment \\
\hline
gs ($3/2^-$)& 0 & 6.1 &  5.7 $\pm$ 0.9\\
\hline
 2.69 MeV $(1/2^-)$ & $\hspace{1.6cm} 2\quad\left\{\begin{array}{l}
                        (\beta=0.1)\\
                        (\beta=0)
                      \end{array}\right.$
  & $\begin{array}{c}
                        0.7 \\
                        5\times 10^{-2}
                      \end{array} $& 1.0 $\pm$ 0.36\\
\hline
\end{tabular}
\caption{\protect\footnotesize Integrated two-neutron differential transfer cross sections associated with the ground state (gs ($3/2^-$))
and with the first excited state (2.69 MeV; $1/2^-$) of $^9$Li in comparison with the data \cite{Tanihata:08}. In the case of the $1/2^-$ state two calculations have been carried out. One making use of the microscopic wavefunction of ref. \cite{Barranco:01} (see Eqs. (\ref{eq1})--(\ref{eq4})). A second one in which it is (arbitrarily) assumed that $\beta=0$ (see Eq. (\ref{eq2})). That is, that the only processes populating the first excited state of $^9$Li are associated with breakup and inelastic channels (see also Fig. \ref{fig3}).}\label{tab2}
\end{footnotesize}
\end{center}
\end{table}

Summing up, through a unified structure--reaction NFT analysis of the experiment of Tanihata et al \citep{Tanihata:08} we are able to conclude that virtual quadrupole vibrations of $^{9}$Li, tailored glue of the halo of $^{11}$Li, in its process of propagating from one partner of the Cooper pair to the other has been caught in the  act by the external pair transfer field produced by the ISAC--2 facility at TRIUMF, forced to become a real final state and to bring this information to the active target detector MAYA. This is a first in the study of pair correlations in nuclei, providing direct information on the central role polarization effects play in nuclear Cooper pair stabilization.

 Discussions with Prof. I. Tanihata and Prof. R. Kanungo are gratefully aknowledged.

%



\end{document}